\documentclass[aps,pre,showpacs,amsfonts,amssymb,amsmath,floatfix,twocolumn,
floats,nobalancelastpage]{revtex4}

\usepackage{graphicx}

\begin{document}          

\title{Algorithmic scalability in globally constrained conservative 
parallel discrete event simulations of asynchronous systems}

\author{A. Kolakowska}
\email{alicjak@bellsouth.net}
\author{M. A. Novotny}
\affiliation{Department of Physics and Astronomy, and the MSU ERC  
P.O. Box 5167, Mississippi State, MS 39762-5167}

\author{G. Korniss}
\affiliation{Department of Physics, Applied Physics, and Astronomy, 
Rensselaer Polytechnic Institute 110 8th Street, Troy, NY 12180-3590}

\date{\today}

\begin{abstract}
We consider parallel simulations for asynchronous systems employing $L$ 
processing elements that are arranged on a ring. Processors communicate 
only among the nearest neighbors and advance their local simulated time 
only if it is guaranteed that this does not violate causality. In 
simulations with no constraints, in the infinite $L$-limit the 
utilization scales (Korniss {\it et al}, PRL {\bf 84}, 2000); but, 
the width of the virtual time horizon diverges (i.e., the 
measurement phase of the algorithm does not scale). In this work, 
we introduce a moving $\Delta$-window global constraint, which modifies 
the algorithm so that the measurement phase scales as well. 
We present results of systematic studies in which the system size 
(i.e., $L$ and the volume load per processor) as well as the 
constraint are varied. The $\Delta$-constraint eliminates the 
extreme fluctuations in the virtual time horizon, provides a 
bound on its width, and controls the average progress 
rate. The width of the $\Delta$-window can serve as a tuning
parameter that, for a given volume load per processor, 
could be adjusted to optimize the utilization so as to maximize 
the efficiency. This result may find numerous applications in 
modeling the evolution of general spatially extended short-range 
interacting systems with asynchronous dynamics, including
dynamic Monte Carlo studies.
\end{abstract}


\maketitle

\section{INTRODUCTION \label{intro}}

Parallel Discrete Event Simulations (PDES) of asynchronous systems 
is a computer science term that stands for parallel simulations 
of complex systems with asynchronous dynamics.  Such spatially extended 
complex interacting systems appear across a wide range of fields in 
natural sciences, and their examples include interacting spin 
systems in material physics, activated processes in chemistry, contact 
processes in stochastic epidemic models, stochastic market models in 
finance, scheduling call arrivals in communication networks, 
and routing problems in internet traffic, to mention just a few 
applications of PDES. Despite active research in this area 
\cite{Fuji90,NF94} very few of the PDES techniques have filtered 
through to the physics community. Even the simplest random-site 
update Monte Carlo schemes \cite{BH97}, where updates correspond to 
Poisson-random discrete events, were long believed to be 
inherently serial (at least in the physics community).  Simulation 
studies of parallel computations for asynchronous distributed 
systems date back more than two decades ago \cite{CM79,CM81}.
However, it was Lubachevsky's work \cite{Lub87,Lub88} on parallel 
simulations of dynamic Ising spin systems which shed a new light 
on this old problem and showed how to efficiently perform conservative 
simulations on a parallel computer. The design of efficient algorithms 
that would allow modeling of asynchronous systems in a 
parallel processing environment is even more important nowadays, 
when parallel architectures have become generally available. 
The architectures of today may consist of several thousands of 
processing elements, but the size of future systems may be of 
the order of hundred thousands \cite{blue02}. Such architectures 
pose new questions of algorithm efficiency and scalability 
in large scale massively parallel processing. We address these 
questions for conservative PDES, using the tools of modern statistical 
physics, in particular, those of non-equilibrium surface 
growth \cite{BS95}.

The difficulties of parallelizing spatially extended asynchronous 
cellular automata arise because in asynchronous systems the discrete 
events are not synchronized by a global clock. For example, in the 
basic dynamic Ising model for ferromagnets discrete spins with 
two states each are placed on a lattice. The discrete events are 
attempted spin flips, where the spin-flip probability at
some site depends on the energy states of the neighboring sites. 
The lattice can be partitioned into a number of sub-lattices, and each 
sub-lattice may be assigned to a different processor. Processing 
elements (PE) attempt a number of randomly chosen spin-flips, and 
communicate with each other in some update attempts (a discrete event). 
Each PE carries its own local virtual time which is advanced 
by every update attempt. The local virtual time on a PE is the 
simulated time at the spins on its sub-lattice. In the conservative 
PDES implementation it is ensured that causality is not violated 
before each PE makes an update attempt. Alternatively, in an 
optimistic PDES implementation, PEs make updates without communicating 
with the neighbors, thus sometimes causing causality errors. The 
optimistic scheme provides a recovery mechanism by undoing the effects 
of all events that have been precessed prematurely. Optimistic 
PDES have been an object of theoretical and simulation studies 
\cite{FK91,Nico91,Stei93,Stei94,Stei96}. The development of 
spatio-temporal correlations and self-organized criticality have been 
recently studied in the optimistic simulations of the dynamics of 
Ising spin systems \cite{SOS01,OSS01}. The conservative scheme has been 
used recently to model magnetization switching \cite{KNR99}, 
ballistic particle deposition \cite{LP96}, and a dynamic 
phase transition in highly anisotropic thin-film ferromagnets 
\cite{KWRM01,KRN02}. These recent applications suggest 
that the conservative scheme should be particularly efficient when 
applied to large systems with short-range interactions.

Early efficiency studies of the conservative scheme \cite{Lub89,Nico93} 
do not identify the mechanism which would ensure the scalability 
of the PDES for an arbitrary system size. Recently, Korniss {\it et al} 
\cite{KTNR00} introduced a new approach that exploits an analogy 
between the virtual time horizon and a fluctuating surface that 
grows in a deposition process. In this picture, the fraction of 
non-idling PEs (the utilization) exactly corresponds to the 
density of local minima in a virtual time surface. They showed that, 
in the case of one spin site per PE, the steady state virtual time 
surface is governed by the Edwards-Wilkinson Hamiltonian, implying 
that the utilization does not vanish for an infinitely large system 
of PEs. Ignoring communication delays and assuming that 
the utilization is equivalent to the efficiency, they concluded 
that the computation phase of short-ranged conservative PDES is 
asymptotically scalable. In general, the utilization should not 
be taken as a sole measure of efficiency in the modeling of 
asynchronous systems. The same analysis of a virtual time 
surface \cite{KTNR00,KNKG02,KNTR01,KNRGT02} 
demonstrated that, in the case of one site per PE, the virtual time 
horizon infinitely roughens in the infinite PE-limit. 
The statistical spread of the virtual time surface severely 
limits an averaging or measurement phase of PDES, 
and divergence leads to severe difficulties with data management. 
Therefore, while the simulation phase (as determined by the 
utilization studies) is asymptotically scalable, the measurement 
phase is not. To ensure the efficiency of the algorithm, solutions 
need to be sought in which both phases of the computation are scalable.

In studies of asynchronous updates in large parallel systems, 
Greenberg {\it et al} \cite{GSS96} proposed a $K$-random connection 
model, where at each time step each PE randomly connects with $K$ 
other PEs in the system. The virtual time horizon for this model 
is short-range correlated and has a finite width in the infinite 
PE-limit. Encouraged by this result, we considered the two 
alternative modifications to the conservative scheme: a random 
connection model \cite{KNGTR} and the moving window 
constraint \cite{KNKG02}. The purpose of these modifications is to 
ensure that the measurement phase of the conservative PDES is scalable.

This paper presents the results of systematic simulation studies of 
conservative asynchronous PDES with the moving window constraint (i.e., 
simulation studies of the simulations). In Section~\ref{PDES} we define 
terminology and we outline both the basic conservative update scheme 
and the constraint update rule that we use in modeling of asynchronous 
PDES. The scheme that we consider is such that the evolution of 
the time horizon is decoupled from the underlying systems. The only 
one assumption that we make about underlying complex systems 
is that they are characterized by short-range interactions. Therefore, 
our analysis is generally valid for a wide spectrum of physics 
applications. Section~\ref{scalability} contains the analysis of 
scalability issues, which is based on analogies between PDES and 
kinetic roughening in non-equilibrium surface growth. In the PDES 
analysis the focus is placed on two major issues: the scaling of the 
simulation phase and the scaling of the measurement phase. In 
Section~\ref{winsec} we present and analyze numerical data that 
were obtained in large scale simulations of an asynchronous 
conservative PDES scheme with a moving window constraint. In our time 
evolution studies, we simultaneously varied the width of the moving 
window and the system size (i.e., the number of processing elements in 
the system as well as the number of volume elements per processing 
element) in search for regularities that would allow general 
conclusions. In Section~\ref{discuss}, we discuss connections between 
the scalability of a constrained conservative scheme and the design of 
highly efficient algorithms for asynchronous systems.

\section{CONSERVATIVE PDES FOR SPATIALLY DECOMPOSABLE CELLULAR AUTOMATA 
\label{PDES}}

We consider an ideal system consisting of $L$ identical PEs, 
arranged on a ring. Each PE has $N_V$ lattice sites (or operation 
volumes) and the algorithm randomly picks one of the $N_V$ sites. 
If the site that is picked is an end site communication with a 
neighboring PE is required, while no communication between 
PEs is required if an interior site is picked. For this system 
a discrete event means an update attempt that takes place 
instantaneously. The state of the system does not change between 
update attempts. Processing elements perform operations concurrently, 
however, update attempts are not synchronized by a global clock. Such 
a system can represent, for example, concurrent operations of 
random spin flipping in a large spatially extended ensemble that can 
be arranged on a regular lattice. In this picture, the ensemble 
is spatially decomposed into $L$ subsystems, each of which carries 
$N_V$ sites. Each subsystem is carried by one PE and the required 
communication is the exchange of information about states of the 
border spins. In the simplest case there is one site per PE, $N_V=1$, 
the system is a closed spin chain, and a spin-flip attempt at 
the $k$-th PE depends on the two nearest-neighbor spins located 
on the $(k-1)$-th and the $(k+1)$-th PEs. The $k$-th PE may not 
perform an update until it receives information from these 
neighboring PEs.

In this conservative PDES scheme, to simulate asynchronous dynamics 
employing $L$ processors, each $k$-th PE generates its own local 
simulated time $\tau_k$ for the next update attempt. Update attempts 
are simulated as independent Poisson-random processes, in which the 
$k$-th local time increment (i.e., the random time interval between 
two successive attempts ) is exponentially distributed with unit 
mean. A processor is allowed to update its local time if it is 
guaranteed not to violate causality. Otherwise, it remains idle. 
The time step $t$ is the index of the simultaneously performed update 
attempt. It corresponds to an integer wall-clock time with each 
processor attempting an update at each value of $t$.

%%%%%%%%%%%%%%%%%%%%%    figure 1    %%%%%%%%%%%%%%%%%%%%%%%%%%%
\begin{figure}[tp]
\includegraphics[width=8.0cm]{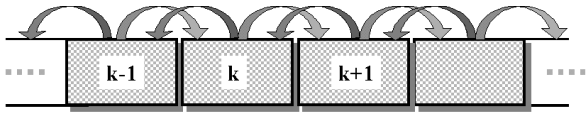}
\caption{\label{fig1} Short range connections in PDES for a linear chain.}
\end{figure}
%%%%%%%%%%%%%%%%%%%%%%%%%%%%%%%%%%%%%%%%%%%%%%%%%%%%%%%%%%%%%%%%%%

The simplest choice for a communication rule between processors, 
which is faithful to the original dynamics of the underlying system, 
is a short-range connection model (Fig.~\ref{fig1}), where, at any 
time step $(t+1)$, the $k$-th PE is allowed to update if its local 
simulated time ${\tau_k} (t)$ is not greater than the local 
simulated times of its two nearest neighbors:
%%%%%%%%%%%%%%%%%%%%%%%%%%%%%%%%%%%%%%%%%%%%%%%%%%%%%%%%%%%%%%%
\begin{equation}
\label{crule}
\tau_k (t) \le \min \left\{ \tau_{k-1} (t), \tau_{k+1} (t) \right\}.
\end{equation}
%%%%%%%%%%%%%%%%%%%%%%%%%%%%%%%%%%%%%%%%%%%%%%%%%%%%%%%%%%%%%%%
The periodicity condition requires communication between the first and 
the last PEs. In effect, this update rule introduces a nearest 
neighbor interaction and corresponding correlations between PEs, 
which is analogous with non-equilibrium surface growth. It was 
shown \cite{KTNR00} that in the case of $N_V=1$ the evolution 
of the virtual time horizon on coarse-grain length scales is 
governed by the Kardar-Parisi-Zhang (KPZ) equation \cite{KPZ86}:
%%%%%%%%%%%%%%%%%%%%%%%%%%%%%%%%%%%%%%%%%%%%%%%%%%%%%%%%%%%%%%%
\begin{equation}
\label{kpz}
\partial _t \tau=\partial _{xx} \tau - \lambda
\left(\partial _x \tau \right)^2 + \eta (x,t),
\end{equation}
%%%%%%%%%%%%%%%%%%%%%%%%%%%%%%%%%%%%%%%%%%%%%%%%%%%%%%%%%%%%%%%
where $x$ is a spatial variable in a continuum limit, the constant 
$\lambda$ is related to the coarse-graining procedure, and $\eta (x,t)$ 
accounts for random fluctuations.

In PDES with a moving window constraint, the communication pattern 
between processing elements remains the short-range connection 
type but the new update rule requires that additionally 
at each $(t+1)$-step the local simulated time of the $k$-th 
processor fits within a window of width $\Delta$ that is defined 
relative to the slowest PE (i.e., the one with the smallest $\tau$) 
at time $t$. Explicitly, the $k$-th PE is allowed the update if 
$\tau _k (t)$ simultaneously satisfies the short-range connection 
condition~(\ref{crule}) and the following window condition:
%%%%%%%%%%%%%%%%%%%%%%%%%%%%%%%%%%%%%%%%%%%%%%%%%%%%%%%%%%%%%%%
\begin{equation}
\label{wrule}
\tau_k (t) \le \Delta +
\min \left\{ \tau_k (t): k=1,\dots,L \right\}.
\end{equation}
%%%%%%%%%%%%%%%%%%%%%%%%%%%%%%%%%%%%%%%%%%%%%%%%%%%%%%%%%%%%%%%
In the computer science community the minimum in Eq.(\ref{wrule}) 
is called the global virtual time 
\cite{FK91,Nico91,Stei93,Stei94}. From this definition it follows that 
the short-range connection model can be viewed as a particular 
case of the original update scheme when the width of the window 
is set to infinity, in which case condition~(\ref{wrule}) is trivially 
satisfied for all times. Thus an infinite window is equivalent to the 
absence of the constraint.

In typical simulations, when the number of volume elements $N_V$ is 
larger than the minimum $N_V=1$, a causality condition~(\ref{crule}) 
is enforced only for the border volumes on each PE. If, at any 
$t$-step, a randomly chosen volume element happens to be from 
the interior, i.e., when all of its immediate neighbors reside on 
one PE, then the PE always executes the update and its local time 
is incremented for the consecutive update attempt: 
$\tau _k (t+1)= \tau _k (t)+ \eta _k (t)$, where $\eta _k$ is 
the $k$-th random time increment that is exponentially distributed, 
randomly chosen independently on each PE and at each parallel step 
$t$. In the constrained simulations, condition~(\ref{wrule}) is enforced 
for any randomly chosen volume element. 

In the conservative update scheme a causality requirement is the main 
mechanism that generates correlations among processing elements. 
In the absence of a causality requirement local simulated times 
would be incremented independently of each other in the fashion of 
random deposition (RD) \cite{BS95}. However, even with this RD update rule, 
imposing the windowing condition~(\ref{wrule}) alone will give rise to 
correlations among processors. 
Note that the RD update rule does not belong to a class of conservative 
update schemes i.e., it cannot faithfully simulate the underlying 
system dynamics. For $\Delta$-constrained RD 
simulations, the speed with which the correlations spread among all 
PEs is determined by the width of the $\Delta$-window.

For a set of $L$ processing elements, a simulated time horizon (STH) 
is defined as a set of $L$ local simulated times at a time 
step $t$. To study the roughening of the STH surface, we monitor the 
surface width $\langle w(t)\rangle$, which is defined in standard fashion 
\cite{BS95} via the variance of the STH:
%%%%%%%%%%%%%%%%%%%%%%%%%%%%%%%%%%%%%%%%%%%%%%%%%%%%%%%%%%%%%%%
\begin{equation}
\label{variance}
\langle w^2(t)\rangle = \left\langle \frac{1}{L} \sum_{k=1}^{L}
\left(\tau _k (t)-\bar{\tau} (t)\right)^2 \right\rangle,
\end{equation}
%%%%%%%%%%%%%%%%%%%%%%%%%%%%%%%%%%%%%%%%%%%%%%%%%%%%%%%%%%%%%%%
where the angle bracket denotes an ensemble average and $\bar{\tau}$ 
denotes the mean virtual time, 
$\bar{\tau}(t)=\frac{1}{L}\sum_{k=1}^{L}\tau _k (t)$. Alternatively, 
the surface width can also be defined as the absolute standard 
deviation $\langle w_a (t) \rangle$ from the mean virtual local time:
%%%%%%%%%%%%%%%%%%%%%%%%%%%%%%%%%%%%%%%%%%%%%%%%%%%%%%%%%%%%%%%
\begin{equation}
\label{absdev}
\langle w_a (t)\rangle= \left\langle \frac{1}{L} \sum_{k=1}^{L}
\left| \tau _k (t)-\bar{\tau} (t) \right| \right\rangle.
\end{equation}
%%%%%%%%%%%%%%%%%%%%%%%%%%%%%%%%%%%%%%%%%%%%%%%%%%%%%%%%%%%%%%%
We use both definitions~(\ref{variance}-\ref{absdev}) in our analysis. To study 
the efficiency of an update process in the system of $L$ processors, 
we define the utilization $\langle u_L(t) \rangle$ as a fraction of processors that 
performed an update at parallel time-step $t$. Throughout the paper we 
consistently use the following notation: the surface width $\langle w(t) \rangle$ is 
an ensemble average of $w(t)=\sqrt{w^2(t)}$ computed at $t$, while 
$\langle w \rangle$ denotes the corresponding steady-state value at 
$t \to \infty$. The subscript ``{\it a}" denotes the width 
computed in accordance to~(\ref{absdev}), while subscripts ``$L$" or 
``$N_V$" (e.g., $\langle w_{L,N_V} \rangle$) indicate the parameter dependance 
of the width computed in accordance to~(\ref{variance}).

\section{SCALABILITY MODELING \label{scalability}}

There are two important aspects of scalability, which should be dealt 
with in studies of algorithm efficiency. Both of them regard the time 
in the system and connect with data management issues. The first 
is the question of whether or not the utilization reaches a constant 
non-zero value in the limit of large system size (when $L$ and $N_V$ 
may arbitrarily vary). In particular, one needs to know if the 
``worst case scenario" of one volume element per PE can produce a 
non-zero utilization in the infinite $L$-limit. A zero 
value of utilization in the infinite system size limit would 
suggest that an algorithm would likely be useless for computationally 
intensive tasks on future generations of massively 
parallel computers, i.e., on systems that contain hundreds of thousands of 
processing elements \cite{blue02}. The second question is the behavior of  
the evolution of the STH, whether or not the 
statistical spread of the STH  saturates in time or scales with the 
system size. A negative answer to the latter question would suggest 
that an algorithm would probably prove impractical in actual 
applications, because the divergence of the STH width implies that, 
for most computationally intensive jobs, the data collection and 
averaging would impose a memory requirement per PE in excess 
of hardware capacities. 

In our scalability studies we exploit existing analogies between the 
time evolution of the STH and kinetic roughening in non-equilibrium 
surface growth; and we use selected results of non-equilibrium surface 
studies \cite{BS95,HZ95,Krug97} in analyzing the stochastic behavior 
of the system under consideration. The conservative short-range 
communication scheme between the system components can be regarded 
as an effective short-range interaction among PEs and treated in a 
similar manner to an interaction among sites of any 
non-equilibrium surface, growing on a regular lattice. For these 
surfaces, the lateral correlation length $\xi$ between sites 
follows the power law $\xi \sim t^{1/z}$, where $z$ is the 
dynamic exponent. For a finite system, $\xi$ cannot grow beyond 
the system size $L$ and it is observed that for times much smaller 
than a cross-over time $t_\times$, $t_\times \sim L^z$, the surface width 
increases in accordance to $t^{\beta}$, where $\beta$ is the 
growth exponent. For times much larger than the cross-over time, 
the surface width saturates and scales as $L^{\alpha}$, 
where $\alpha$ is the roughness exponent. The exponents satisfy 
the scaling relation: $z\beta = \alpha$. The values of the exponents 
are independent of the details of the system and of the nature of the 
interactions between sites. Their values can be determined from the 
corresponding stochastic growth equation, which defines the 
universality class. We observed that the simulated time horizon shows 
kinetic roughening and the typical scaling behavior \cite{KNKG02}:
%%%%%%%%%%%%%%%%%%%%%%%%%%%%%%%%%%%%%%%%%%%%%%%%%%%%%%%%%%%%%%%
\begin{eqnarray}
%\label{scale}
\langle w_L^2 (t) \rangle & \sim &
 t^{2 \beta}  \quad  \text{for $t \ll t_\times$,} \label{scale1} \\
\langle w_L^2 (t) \rangle & \sim &
 L^{2 \alpha}  \quad  \text{for $t \gg t_\times$.} \label{scale2}
\end{eqnarray}
%%%%%%%%%%%%%%%%%%%%%%%%%%%%%%%%%%%%%%%%%%%%%%%%%%%%%%%%%%%%%%%
It was demonstrated in \cite{KTNR00} that in the case of one site per 
PE (i.e., $N_V=1$) the time evolution of the STH in the short-range 
connection model~(\ref{crule}) follows the KPZ equation~(\ref{kpz}) and 
direct simulations confirmed that the scaling exponents 
in Eqs.~(\ref{scale1}-\ref{scale2}) have values consistent with the 
KPZ universality class ($\alpha =1/2$ and $\beta = 1/3$).

\subsection{Steady-state scaling for utilization \label{utilization}}

%%%%%%%%%%%%%%%%%%%%    figure 2     %%%%%%%%%%%%%%%%%%%%%%%%%%%%
\begin{figure}[bp]
\includegraphics[width=8.0cm]{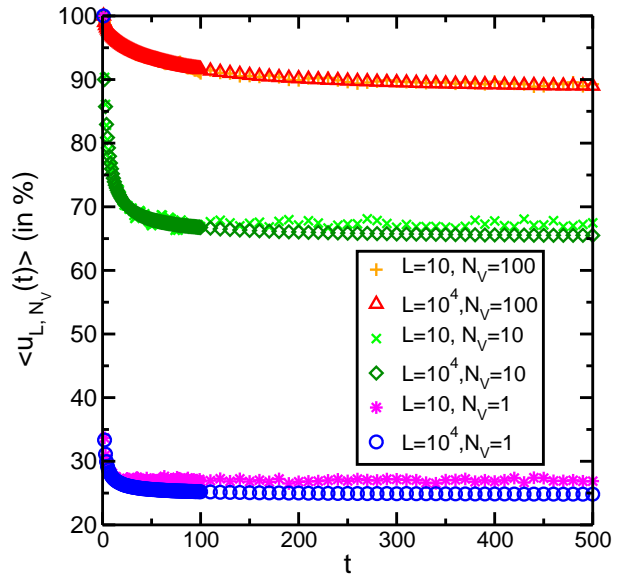}
\caption{\label{fig2} Unconstrained PDES: Time evolution of the mean
utilization $\langle u(t) \rangle$ (averaged over $N=1024$ independent
random trials) for various system sizes: $L=10$ and $10^4$ and $N_V =1, 10$ and $100$.}
\end{figure}
%%%%%%%%%%%%%%%%%%%%%%%%%%%%%%%%%%%%%%%%%%%%%%%%%%%%%%%%%%%%%%%%%%

As the time index advances the utilization falls from its initial 
maximal value at $t=0$. Figure~\ref{fig2} presents the time 
evolution of the utilization for various system sizes in 
the basic PDES with short-range connections with the infinite 
$\Delta$-window. For each of the system size, the utilization 
attains a steady-state, characterized by a non-zero value in 
the infinite $t$-limit. This qualitative result is also true for the 
simulations in two and three dimensions, when an individual PE is 
allowed to connect with four and six immediate neighbors, 
respectively \cite{KNTR01}. Such a non-zero steady-state value is 
characteristic for the KPZ universality class and can be expressed 
by the  Krug and Meakin \cite{KM90} relation for generic KPZ-like 
processes:
%%%%%%%%%%%%%%%%%%%%%%%%%%%%%%%%%%%%%%%%%%%%%%%%%%%%%%%%%%%%%%%
\begin{equation}
\label{krug}
\lim_{t \to \infty} \langle u_L (t) \rangle \approx \langle u_ \infty \rangle + 
\frac{const.}{L^{2(1-\alpha)}},
\end{equation}
%%%%%%%%%%%%%%%%%%%%%%%%%%%%%%%%%%%%%%%%%%%%%%%%%%%%%%%%%%%%%%%
where $\langle u_ \infty \rangle$ denotes the utilization in the infinite $L$-limit. 
Toroczkai {\it et al} \cite{TKDZ00} showed that the basic 
conservative PDES with one site per PE satisfies relation~(\ref{krug}), 
and they used it to extrapolate their utilization data to large $L$. 
Their value for the utilization in the infinite PE-limit is 
$\langle u_ \infty \rangle=24.6461(7)\%$ \cite{KTNR00,KNKG02}. This finding 
demonstrates that the simulation part of the algorithm is scalable 
in the case of the {\it 1-d} conservative PDES with the minimal number 
of volume elements per PE. Explicitly, this means that even in the 
worst case scenario, it is possible to run simulations arbitrarily 
long with a non-zero average rate of progress. In the case of {\it 2-d} 
and {\it 3-d} PDES, the roughness exponents are $\alpha =0.2-0.4$ 
(in {\it 2-d}) and $\alpha =0.08-0.3$ (in {\it 3-d}) \cite{KNTR01}, 
and the $N_V =1$ steady-state utilization can be estimated as 
$\langle u_ \infty \rangle \cong 12\%$ and 
$\langle u_ \infty \rangle \cong 7.5\%$, respectively \cite{KNTR01}.

\subsection{The evolution of the simulated time horizon \label{sth}}

\begin{figure}[tp!]
\includegraphics[width=8.0cm]{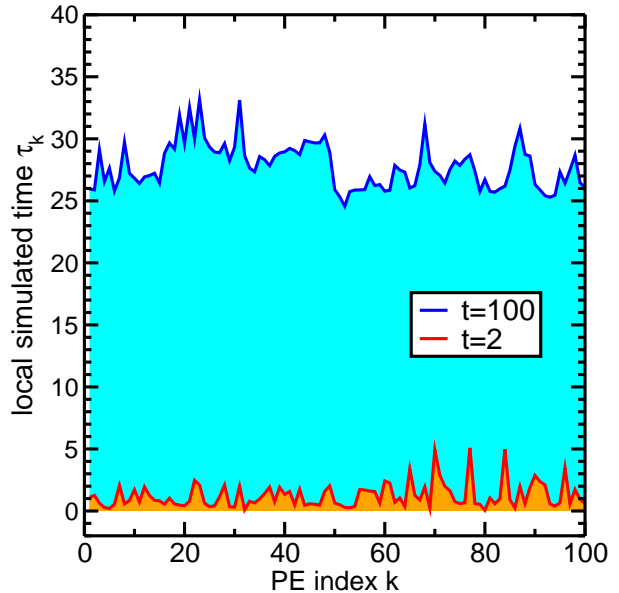}
\caption{\label{fig3} Unconstrained PDES: Time evolution of the time horizon
for $L=100$ PEs and $N_V=1$ sites per PE. The lower surface
is a snapshot at $t=2$, the upper surface is a snapshot at $t=100$.
For $L=100$ the crossover time is $t_\times \approx 3700$.}
\end{figure}

\begin{figure*}[tp!]
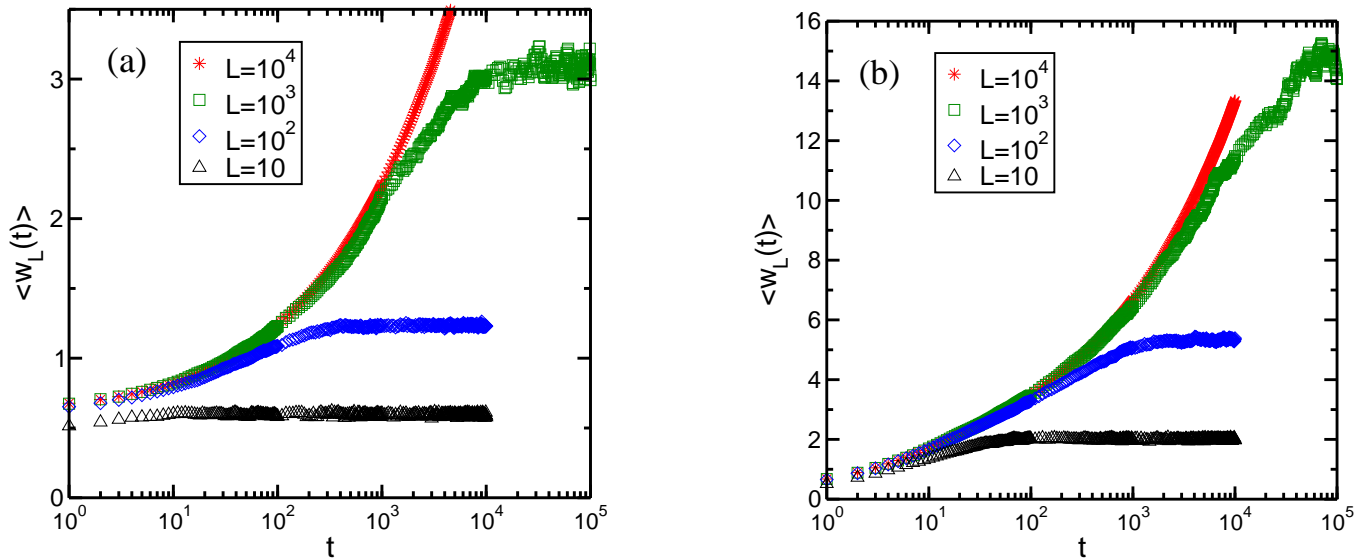

\includegraphics[width=8.0cm]{pre02-cf4a.eps}
\hfill
\includegraphics[width=8.0cm]{pre02-cf4b.eps}
\caption{\label{fig4} Unconstrained PDES: Time evolution of the mean surface
width $\langle w(t) \rangle$ of the STH (averaged over $N=1024$ independent
random trials) for various number $L$ of PEs, in simulations
with: (a) $N_V=1$ site per PE; (b) $N_V=10$ sites per PE.
Since the plateau has been reached for $L=10$ and $L=100$, the times $t$
larger than $10^4$ are not shown. For $L=10^4$, the plateau is
reached for $t$ larger than $10^6$.}
\end{figure*}

The unconstrained PDES are characterized by an infinite roughening 
of the STH surface in the limit of infinite system size. 
Figure~\ref{fig3} presents a typical time evolution of a surface 
generated by this basic update scheme for $N_V=1$ and $L=100$. 
As the time index advances, the surface grows and the statistical 
spread of its interface increases in accordance with 
Eqs.~(\ref{scale1}-\ref{scale2}). Figure~\ref{fig4} shows the time evolution of 
the surface width for a few selected system sizes. For a fixed system 
size the width follows relations~(\ref{scale1}-\ref{scale2}): after the 
initial growth phase, the surface saturates and its width reaches the 
plateau value. By comparing the widths for $N_V=1$ (Fig.~\ref{fig4}a) 
to those for $N_V=10$ (Fig.~\ref{fig4}b), one can see that for a 
fixed $L$-number of PEs, increasing the number of sites per PE 
shifts the cross-over time $t_\times$ to later values and increases 
the value of the width at the plateau. This is expected since a 
larger value of $N_V$ yields a larger cumulative value for the local 
time increment between two successive communications with neighboring 
PEs. In the case of $N_V=1$, the width of the STH approaches a finite 
constant for a finite $L$-number of PEs; however this constant 
diverges in the infinite $L$-limit in the power law fashion:
%%%%%%%%%%%%%%%%%%%%%%%%%%%%%%%%%%%%%%%%%%%%%%%%%%%%%%%%%%%%%%%
\begin{equation}
\label{diverge}
\langle w^2 \rangle \sim L^{2 \alpha},
\end{equation}
%%%%%%%%%%%%%%%%%%%%%%%%%%%%%%%%%%%%%%%%%%%%%%%%%%%%%%%%%%%%%%%
which gives the linear divergence of the surface variance for the 
KPZ universality class. The same holds for the extreme fluctuations 
above and below the mean simulated time \cite{KTNR00}. This finding 
is also valid in the case when each PE carries a block of sites. 
The above scaling behavior creates difficulties when intermittent 
data on each PE have to be stored until all PEs reach the simulated 
time instant at which statistics collection is performed (e.g., 
a simple averaging over the full physical application). The diverging 
spread of the time horizon implies a diverging storage need for this 
purpose on every PE. Thus, the diverging width of the STH means that 
the memory requirement per processing element, for temporary data 
storage, diverges as the number of processing elements gets 
arbitrarily large. Therefore, the measurement phase of an 
algorithm that follows the basic conservative update scheme is 
asymptotically not scalable. In actual applications, the programmer 
must seek some means of constraining the infinite roughening of the 
STH or must impose some global synchronization on the system of PEs.

%%%%%%%%%%%%%%%%%%%%    figure 3     %%%%%%%%%%%%%%%%%%%%%%%%%%%%
%%%%%%%%%%%%%%%%%%%%%%%%%%%%%%%%%%%%%%%%%%%%%%%%%%%%%%%%%%%%%%%%%%

%%%%%%%%%%%%%%%%%%%%    figure 4     %%%%%%%%%%%%%%%%%%%%%%%%%%%%
%%%%%%%%%%%%%%%%%%%%%%%%%%%%%%%%%%%%%%%%%%%%%%%%%%%%%%%%%%%%%%%%%%

Our and Lubachevsky's earlier studies show that to make the 
conservative scheme efficient, one must use a large number of 
volume elements $N_V$. It is expected that an increase in $N_V$ 
will modify the growth phase of the STH. In the case of large $N_V$, 
the initial growth phase (for $0<t<t_1$) should be characterized by 
$\beta=1/2$, typical for the RD universality class. Then, after the 
first cross-over time $t_1$ (for $t_1<t<t_2$, where $t_2$ is 
the saturation time) the growth should be characterized by 
$\beta=1/3$, typical for the KPZ universality class. In this way, 
making the simulation phase more efficient (by increasing $N_V$) 
will speed up the initial growth. Thus, the state savings, which 
are traditionally associated with optimistic schemes, 
are disadvantageous in conservative schemes.

\section{CONSERVATIVE PDES WITH THE MOVING WINDOW CONSTRAINT 
\label{winsec}}

A standard way of controlling the growth of the STH is to impose a 
constraint on its width in the spirit of parallel simulations of 
asynchronous cellular automata that was proposed by Lubachevsky 
\cite{Lub87,Lub88}. A straightforward application of this idea is 
the $\Delta$-constrained update scheme which demands that at each 
update attempt a PE can perform an update only if its value 
of $\tau_k$ is within the window. The effect of condition~(\ref{wrule}) 
is that fast PEs are forced to postpone their updates until
slower PEs catch up. In the simplified version studied here, the assigned distance 
apart is measured in terms of the processor local time that 
is compared to the global minimum virtual time. Since at each $t$ 
the global minimum of the STH changes its location, so does 
the window for the update. In this sense a moving window constraint 
can be considered as an implicit rule that induces global 
synchronization, in which each PE connects with the slowest PE. 
From the implementation point of view, the most important questions 
are the scalability issues  for realistic systems, where each PE 
may carry an arbitrary number of sites, because mainly these issues 
will determine the efficiency of the algorithm in actual applications.

\subsection{Simulation phase \label{simulation}}

In the $\Delta$-constrained PDES, simulations reach a steady state 
for an arbitrary system size in a similar fashion as in the basic 
short-range connection model, illustrated in Fig.~\ref{fig2}. 
In general, for any $\Delta$-value, when $L$ is fixed the steady state 
value of the utilization gets larger as  $N_V$ gets larger; and, 
when $N_V$ is fixed it gets smaller as $L$ is increased. This 
behavior reflects the strength of the correlations between PEs 
which arise due to the update rule~(\ref{crule}). Namely, for fixed $L$, 
the frequency of an update per PE increases as $N_V$ increases 
because condition~(\ref{crule}) does not need to be verified for 
the internal sites and the probability of randomly choosing a border 
site is $1/N_V$. Therefore, in this case, correlations that 
arise due to the short-range connections between PEs weaken when 
$N_V$ is increasing. In the infinite $N_V$-limit these 
correlations become negligible and the process of incrementing local 
simulated times resembles random deposition on the {\it 1-d} lattice 
of size $L$. Thus, the RD-limit is equivalent to the infinite 
$N_V$-limit of PDES. 

\begin{figure*}[tp!]
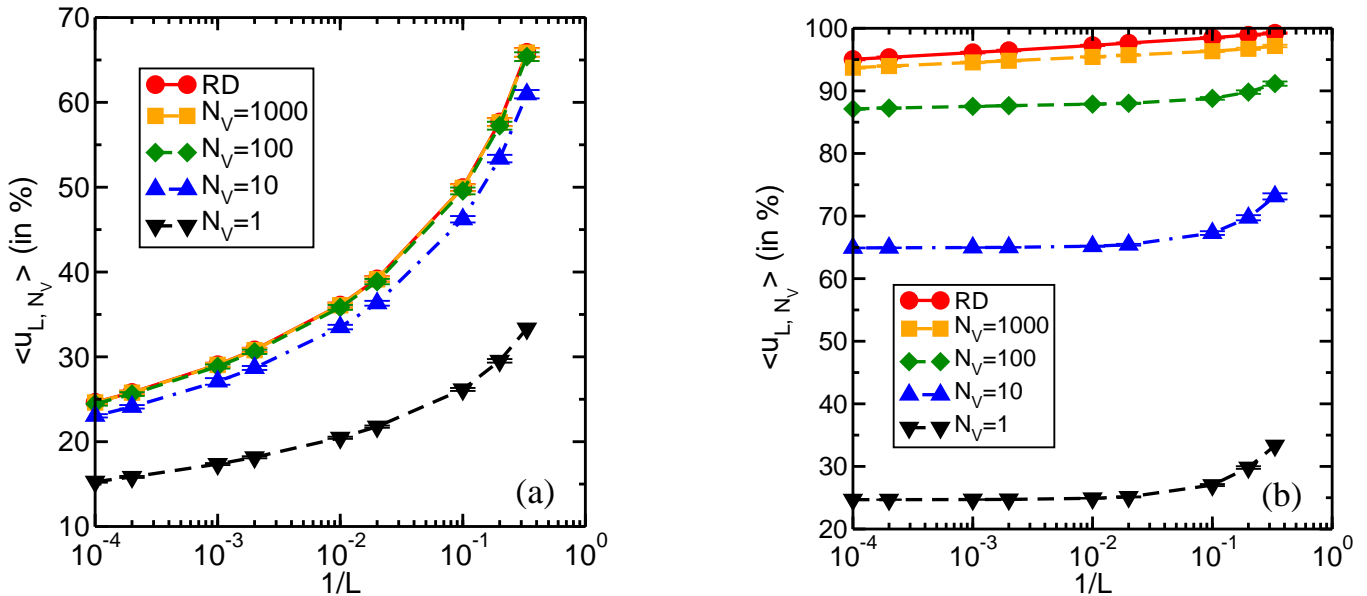

\includegraphics[width=8.0cm]{pre02-cf5a.eps}
\hfill
\includegraphics[width=8.0cm]{pre02-cf5b.eps}
\caption{\label{fig5} Mean steady state utilization
$\langle u \rangle $ in constrained PDES as
a function of the system size for the $\Delta$-window size:
(a) $\Delta =10$; (b) $\Delta =100$. $L$ is the number of PEs. When
the number $N_V$ of sites per PE is increased the curves converge
towards the RD limit.}
\end{figure*}

The mean steady state utilization $\langle u_{L,N_V} \rangle$ as a function of the 
system size is presented in Fig.~\ref{fig5}. When the number 
$N_V$ of sites per PE is increased the curves converge towards 
the RD limit. With a narrow $\Delta$-window (Fig.~\ref{fig5}a) 
the RD limit is approached fairly quickly (with $N_V=100$ for 
$\Delta=10$), while with a wide $\Delta$-window (Fig.~\ref{fig5}b) 
the RD limit is approached more slowly. For an infinite 
$\Delta$-window the RD limit is $\langle u_{L,\infty} \rangle=100\%$, which is the 
effect of no-correlations in the system in this limiting case. 
Obviously, $\langle u_L \rangle=1/L \times 100\%$ when $\Delta =0$, because in 
this case only one PE is allowed to make an update. The RD curves in 
Fig.~\ref{fig5} display the steady state utilization for simulations 
that are governed only by the update rule~(\ref{wrule}) alone, i.e., in 
the absence of other communications between processing elements. The 
fall off in the RD utilization values with an increase in the number 
of PEs, indicates the strength of correlations between PEs, which 
exclusively results from imposing the $\Delta$-window constraint. 
When all three parameters, $\Delta$, $L$ and $N_V$, are allowed to 
vary in conservative PDES, the value of the utilization is mostly 
determined by the width of the moving window. The choice of a very 
narrow $\Delta$-window severely suppresses the average progress rate.

%%%%%%%%%%%%%%%%%%%%    figure 5     %%%%%%%%%%%%%%%%%%%%%%%%%%%%
%%%%%%%%%%%%%%%%%%%%%%%%%%%%%%%%%%%%%%%%%%%%%%%%%%%%%%%%%%%%%%%%%%

To determine a scaling relation for the steady state utilization in 
the infinite PE-limit, we analyzed the mean steady state 
utilization $\langle u_L \rangle$ as a function of $1/L$ for several values of 
$\Delta$ and $N_V$, in each case performing a standard rational 
function interpolation \cite{PTVF92} of the simulation data:
%%%%%%%%%%%%%%%%%%%%%%%%%%%%%%%%%%%%%%%%%%%%%%%%%%%%%%%%%
\begin{eqnarray}
\label{approx}
\langle u_L \rangle &=& \frac{a_0+\sum_{k=1}^{K_n}a_k 
\left( \frac{1}{L} \right)^k}{1+
\sum_{k=1}^{K_d} b_k \left( \frac{1}{L}\right)^k}
 \\
 &=& \frac{a_0}{1+\sum_{k=1}^{K_d} b_k
\left( \frac{1}{L} \right) ^k}
+\frac{a_1}{L} \frac{1+\sum_{k=1}^{K_n-1} \frac{a_{k+1}}{a_1}
\left( \frac{1}{L} \right)^k }{1+\sum_{k=1}^{K_d} 
b_k \left( \frac{1}{L}\right)^k}, \nonumber
\end{eqnarray}
%%%%%%%%%%%%%%%%%%%%%%%%%%%%%%%%%%%%%%%%%%%%%%%%%%%%%%%%%
where the polynomial degrees $K_n$ and $K_d$ were varied to determine 
the best set of the interpolation coefficients. Then, knowing the 
leading coefficients $a_0$ and $a_1$, we extrapolated the utilization 
values to $L=\infty$. In the infinite $L$-limit the leading term in 
Eq.~(\ref{approx}) is $\langle u_{\infty} \rangle \equiv a_0$ and we obtain the 
following scaling relation:
%%%%%%%%%%%%%%%%%%%%%%%%%%%%%%%%%%%%%%%%%%%%%%%%%%%%%%%%%
\begin{equation}
\label{myscale}
\lim_{L \to \infty} \langle u_L \rangle =
\langle u_\infty \rangle + \frac{const.}{L}.
\end{equation}
%%%%%%%%%%%%%%%%%%%%%%%%%%%%%%%%%%%%%%%%%%%%%%%%%%%%%%%%%

\begin{figure}[tp]
\includegraphics[width=8.0cm]{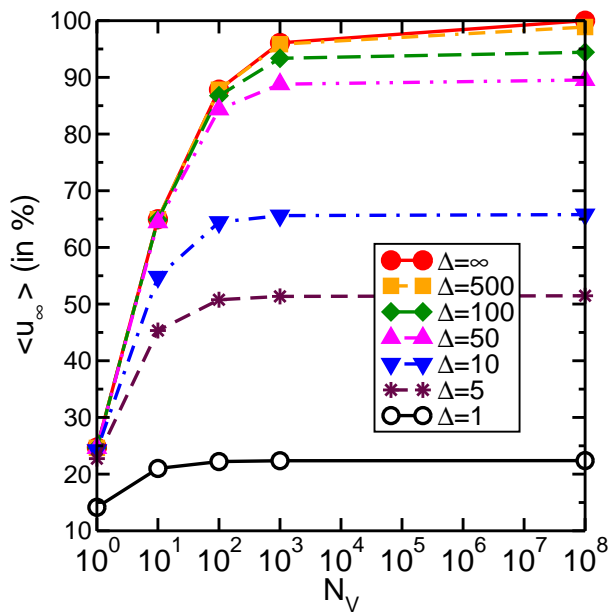}
\caption{\label{fig6} Mean utilization
$\langle u_{\infty} \rangle$ in the limit of $L \to \infty$
as a function of the number of volume elements $N_V$ and the
$\Delta$-window size. Data points for $N_V=10^8$ present the
constrained RD simulations. Symbols represent the simulation data. 
The lines are guides for the eyes.}
\end{figure}

The mean utilization $\langle u_\infty \rangle$ in the infinite $L$-limit, as a 
function of $N_V$ and the $\Delta$-window size, is presented in 
Fig.~\ref{fig6}. Data points for $N_V=10^8$ represent extrapolated 
values for the $\Delta$-constrained RD simulations. It can be 
observed that in the infinite $L$-limit, as well as at each update 
attempt and in the saturation limit, the utilization is strongly 
affected by the value of $\Delta$. A narrow $\Delta$-window can 
significantly slow down the system because a significant number of 
PEs (that otherwise would perform an update) may be constrained to 
wait for the slowest ones to catch up. This effect is particularly 
noticeable when the number $N_V$ of sites per PE becomes large. 
For example, for $N_V=100$, when the $\Delta$-window is narrowed 
to $\Delta =1$, the utilization may drop by as much as $65\%$ from 
its value at $\Delta =100$. When $\Delta =0$, $\langle u_\infty \rangle=0$ for 
any $N_V \ge 1$.

%%%%%%%%%%%%%%%%%%%%    figure 6     %%%%%%%%%%%%%%%%%%%%%%%%%%%%
%%%%%%%%%%%%%%%%%%%%%%%%%%%%%%%%%%%%%%%%%%%%%%%%%%%%%%%%%%%%%%%%%%

The standard $\%$-error in our simulation data for the utilization 
at each $t$-step does not exceed $1\%$ when configurational averages 
are extracted from $N=1024$ independent random trials, except for 
the data obtained with the infinite window, which are within a $2\%$ 
error bar (due to a smaller $N$). We estimate that our values for the 
steady state utilization in the infinite $L$-limit are well within 
a $2\%$ relative uncertainty. The utilization data that are presented 
in Fig.~\ref{fig6} can be fitted to the approximate formula:
%%%%%%%%%%%%%%%%%%%%%%%%%%%%%%%%%%%%%%%%%%%%%%%%%%%%%%%%%
\begin{equation}
\label{fit}
u(N_V,\Delta)=u_{RD}(\Delta) \times u_{KPZ}
({N_V})^{p(\Delta,N_V)},
\end{equation}
%%%%%%%%%%%%%%%%%%%%%%%%%%%%%%%%%%%%%%%%%%%%%%%%%%%%%%%%%
where the first factor approximates the utilization curve in the RD 
limit, $u_{RD}(\Delta)=\lim_{N_V \to \infty} u(N_V,\Delta)$. The 
base in the second factor approximates the utilization curve in the 
infinite $\Delta$-limit, 
$u_{KPZ}(N_V)= \lim_{\Delta \to \infty} u(N_V,\Delta)$. Justification 
and the details of the fit~(\ref{fit}) are outlined in the Appendix. Here, 
$u(N_V,\Delta)$ denotes an approximate value of 
$\langle u_{N_V,\Delta} \rangle$.

A mean-field type argument can also be used to derive an approximate 
formula for $u_{KPZ}(N_V)$: 
%%%%%%%%%%%%%%%%%%%%%%%%%%%%%%%%%%%%%%%%%%%%%%%%%%%%%%%%%
\begin{equation}
\label{man1}
\frac{1}{u_{KPZ}(N_V)} -1= \left( \delta - \frac{2}{N_V} \right) p_w ,
\end{equation}
%%%%%%%%%%%%%%%%%%%%%%%%%%%%%%%%%%%%%%%%%%%%%%%%%%%%%%%%%
where $\delta$ depends on $N_V$ and is the average number of steps 
a PE waits given that it has to inquire about the virtual time on 
its neighboring PEs, when the simulations reach the steady state in 
the system of the infinite number of PEs. Equation~(\ref{man1}) 
is valid for $N_V \ge 3$, where the mean-field approximation of 
replacing the average of a function with the function of the 
averages has been used. In justification for Eq.~(\ref{man1}) 
we assume that a neighboring PE has a virtual time which lags behind 
that of the checking PE, hence requiring the checking PE to wait. Let 
the total number of times on average a PE advances be 
$n_{tot}=n_{OK}+n_w$, where $n_{OK}$ is the number of times it does 
not have to wait and $n_w$ is the number of times it has to 
wait for its neighboring PE. Then, in a mean-field spirit: 
$u_{KPZ}(N_V)=n_{tot}/(n_{OK}+ \delta n_w)=1/(p_{OK}
+ \delta p_w)$, where 
$p_{OK}=n_{OK}/n_{tot}$ and $p_w=n_w/n_{tot}$. Probability 
$p_{OK}$ is the probability of not having to wait when either an 
interior site or a border site is chosen: 
$p_{OK}=(N_V -2)/N_V + (1 - p_w) \left( 2/N_V \right)$, where  
$p_w$ is the probability 
of waiting when either of the border sites is chosen. 
Combining $u_{KPZ}$ and $p_{OK}$ gives Eq.~(\ref{man1}).

Similar arguments can be used to derive an approximate formula in the 
limit of large $\Delta$:
%%%%%%%%%%%%%%%%%%%%%%%%%%%%%%%%%%%%%%%%%%%%%%%%%%%%%%%%%
\begin{equation}
\label{man2}
\frac{1}{u(\Delta,N_V)}-1= \left( \delta - \frac{2}{N_V} \right) p_w + 
\left( \kappa -1 + \frac{2}{N_V} p_w \right) p_{\Delta},
\end{equation}
%%%%%%%%%%%%%%%%%%%%%%%%%%%%%%%%%%%%%%%%%%%%%%%%%%%%%%%%%
where $\kappa$ depends on both $N_V$ and $\Delta$, and is the average 
number of steps a PE waits given that it does not have to wait for a 
neighboring PE but has to wait because of the $\Delta$-window 
constraint. The meaning of $\delta$ and $p_w$ is as in Eq.~(\ref{man1}). 
Let $n_w$ be the number of times the PE waits given that a border 
site has been chosen, and $n_{\Delta}$ be the number of times a PE 
waits because the $\Delta$-condition has been not satisfied either 
at the border or at the interior site. The corresponding probability 
$p_\Delta$ is the probability of waiting because the $\Delta$-condition 
is not satisfied. In justification for 
Eq.~(\ref{man2}) we assume that in the limit of large $\Delta$, 
the events of violating the window condition at the border are 
almost disjoint from the events of violating the nearest-neighbor 
update condition. With this assumption, no matter which is done, 
one cycle will be used to update the configuration, so the total 
number of updates is $n_{tot}=n_{OK}+n_w+n_{\Delta}$, while 
the number of cycles taken on average is 
$n_{OK}+ \delta n_w + \kappa n_{\Delta}$. 
Defining the probabilities as above, 
yields the approximate relation (\ref{man2}). Additional 
approximations can be made by assuming uniformly distributed waiting 
times. Note that for fixed $N_V$ and $\Delta$, both $\delta$ and 
$\kappa$ can be measured independently of the utilization, thereby testing 
the mean-field spirit of the calculation.

\subsection{Measurement phase \label{measure}}

\begin{figure}[tp]
\includegraphics[width=8.0cm]{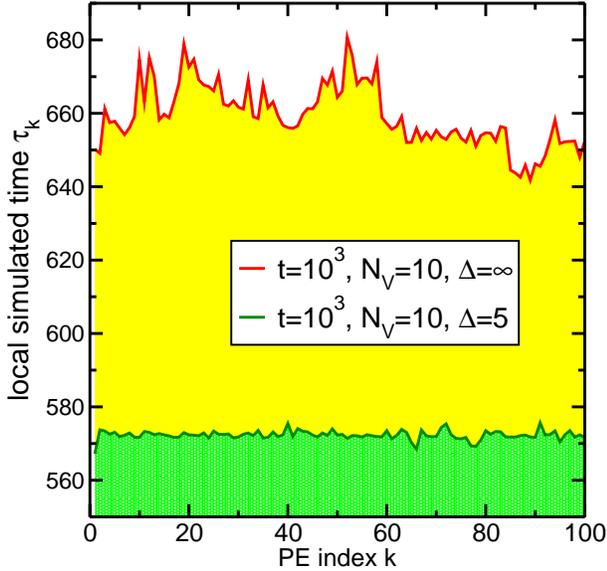}
\caption{\label{fig7} The roughening of the STH. For $\Delta = \infty$ 
(the upper surface), the crossover time is $t_\times \approx 4000$, 
and for $\Delta =5$ (the lower surface) 
the width begins to saturate at $t_p \approx 40$.}
\end{figure}

Direct simulations show that the $\Delta$-constrained width of the 
STH is bounded: its absolute spread remains within the $\Delta$-value 
for an arbitrary system size. This result should be expected since 
the update rule~(\ref{wrule}) implies that independently of the system 
size, at each update attempt, the absolute deviation from the 
minimum cannot take on values much larger than $\Delta$ (if it does, 
the update does not happen). Thus, $w_a$ as well as $w$ may not 
exceed $\Delta$. The surface of the STH is effectively smoothed 
at each update attempt. Figure~\ref{fig7} shows the difference in 
roughening for two surfaces after $t=1000$ steps: the upper 
surface  is obtained in simulations without the $\Delta$-constraint, 
while the lower surface is obtained with $\Delta =5$. 

%%%%%%%%%%%%%%%%%%%%    figure 7     %%%%%%%%%%%%%%%%%%%%%%%%%%%%
%%%%%%%%%%%%%%%%%%%%%%%%%%%%%%%%%%%%%%%%%%%%%%%%%%%%%%%%%%%%%%%%%%

\begin{figure}[tp]
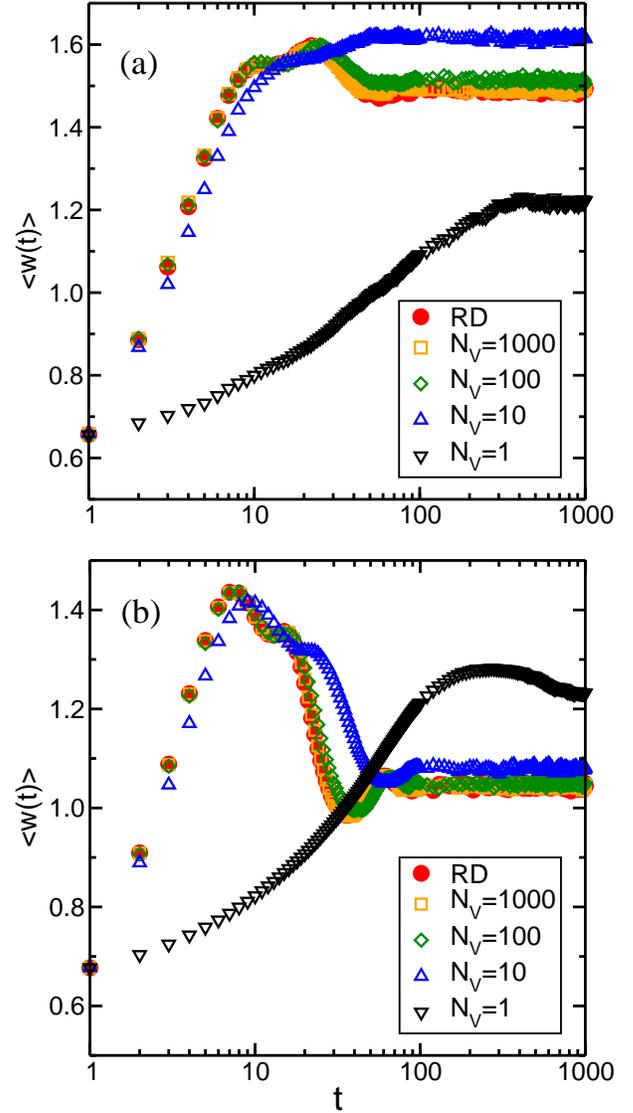

\includegraphics[width=8.0cm]{pre02-cf8a.eps}
\includegraphics[width=8.0cm]{pre02-cf8b.eps}
\caption{\label{fig8} The time evolution of the mean STH surface width
$\langle w(t) \rangle$ (averaged over $N=1024$ independent random trials) in
PDES with $\Delta =10$, for: (a) $L=100$; (b) $L=1000$. 
The curves are plotted for several $N_V$. Plots of 
$\langle w_a(t) \rangle$ look exactly the same.}
\end{figure}

\begin{figure*}[tp!]
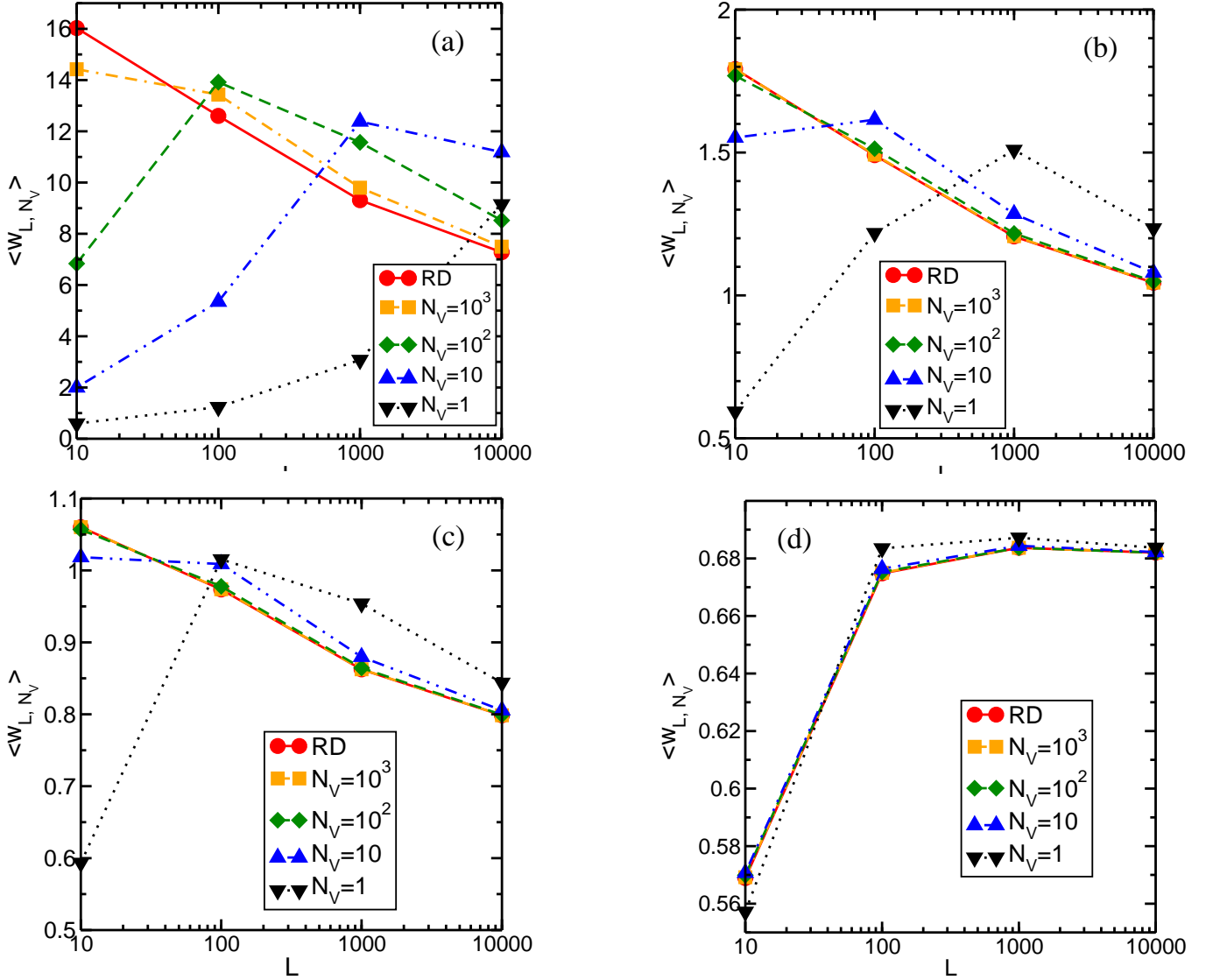

\includegraphics[width=8.0cm]{pre02-cf9a.eps}
\hfill
\includegraphics[width=8.0cm]{pre02-cf9b.eps}
\vfill
\includegraphics[width=8.0cm]{pre02-cf9c.eps}
\hfill
\includegraphics[width=8.0cm]{pre02-cf9d.eps}
\caption{\label{fig9} PDES with the $\Delta$-window constraint: The steady
state surface width $\langle w \rangle$ of the STH as a function of the system size.
(a) $\Delta =100$; (b) $\Delta =10$; (c) $\Delta =5$; and,
(d) $\Delta =1$. The curves are plotted for several values of
volume elements $N_V$. The lines are guides for the eyes.}
\end{figure*}

The $\Delta$-constrained time surfaces exhibit the initial growth and 
the saturation at later times, similar to Fig.~\ref{fig4}. However, 
a detailed analysis of the time evolution of the surface width 
suggests that, in general, they do not belong to the KPZ universality 
class unlike surfaces generated with $\Delta = \infty$.  
Figure~\ref{fig8} presents a typical behavior of the width 
for $\Delta =10$. In general, the transition to the saturated state 
takes place over a time interval (a ``bump" in Fig.~\ref{fig8}), whose 
length and position depends mainly on the $\Delta$-value and cannot 
be characterized by a single crossover time. In the initial growth 
phase for $t \ll t_p$ ($t_p$ marks the beginning of the plateau, 
Fig.~\ref{fig8}), the surface width scales as $t^{\beta}$, i.e., 
for a fixed $\Delta$ and $N_V$, the growth is characterized by one 
value of exponent $\beta$ for any $L$. In general, surfaces generated 
with various values of parameters $\Delta$ and $N_V$ are characterized 
by various effective values of the growth exponent $\beta$. When 
$\Delta =\infty$, $\beta$ values are between the KPZ value of $1/3$ 
(for $N_V=1$) and the RD value of $1/2$ (for $N_V=\infty$) for small 
and intermediate $N_V$ and $L$. In the saturated phase, 
$t_p \ll t$, for a fixed value of $\Delta$, the surface width 
$\langle w_{L,N_V}(t) \rangle$ decreases with the system size, as can be observed 
from Fig.~\ref{fig8}. The saturated surface width as a function of 
the system size is plotted for $\Delta =100, 10, 5, 1$ in 
Fig.~\ref{fig9}. It can be seen that increasing the number of PEs and 
the number of sites per PE does not result in infinite roughening of 
the STH.

%%%%%%%%%%%%%%%%%%%%    figure 8     %%%%%%%%%%%%%%%%%%%%%%%%%%%%
%%%%%%%%%%%%%%%%%%%%%%%%%%%%%%%%%%%%%%%%%%%%%%%%%%%%%%%%%%%%%%%%%%

%%%%%%%%%%%%%%%%%%%%    figure 9     %%%%%%%%%%%%%%%%%%%%%%%%%%%%
%%%%%%%%%%%%%%%%%%%%%%%%%%%%%%%%%%%%%%%%%%%%%%%%%%%%%%%%%%%%%%%%%%

The STH produced in the RD simulations with the infinite 
$\Delta$-window (in other words, in PDES with no communication between 
PEs) is characterized by a surface that is not self-affine \cite{BS95}. 
Nonetheless, the presence of a finite $\Delta$-window constraint in the 
RD simulations forces the STH surface to saturate (Fig.~\ref{fig8}). 
Therefore, this type of PDES no longer belongs to the RD universality 
class, characterized by $\beta =0.5$ and $\alpha = \infty$. In 
the $\Delta$-constrained PDES, the $\Delta$-constrained RD surface 
is the limiting case when the number of sites per PE grows to infinity. 

\begin{figure}[tp!]
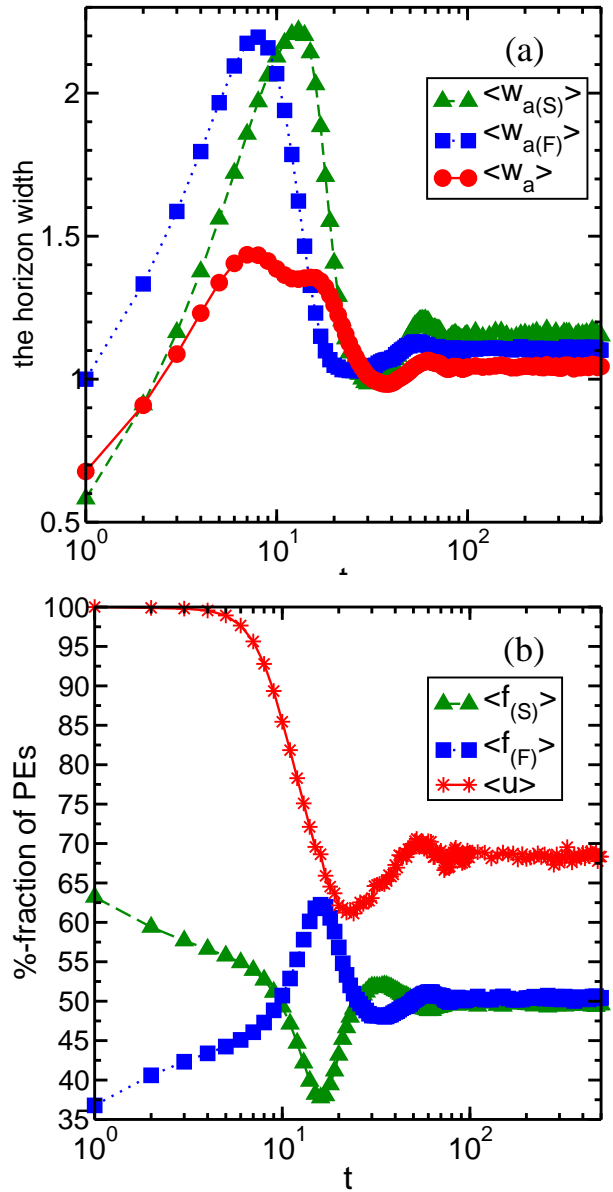

\includegraphics[width=8.0cm]{pre02-cf10a.eps}
\vfill
\includegraphics[width=8.0cm]{pre02-cf10b.eps}
\caption{\label{fig10} PDES with  $\Delta=10$, $N_V=10^3$ and $L=10^4$:
(a) Time evolution of the surface widths; (b) Time evolution of
$\%$-fractional contributions to the surface widths. Subscripts (S)
and (F) denote a fraction of processors in the slow and in the fast
group, respectively; and $u$ is the utilization. Configurational
averages were taken over $N=1024$ independent random trials. The
lines are guides for the eyes.}
\end{figure}

An interesting feature in the surface width evolution graphs 
(Fig.~\ref{fig8} and Fig.~\ref{fig10}) is the presence of a maximum that 
marks the end of the growth phase. Its double peak structure can be 
explained, both quantitatively and qualitatively, in terms of simplex 
geometry \cite{Alfs71}. In the set of $L$ processing elements we 
distinguish between slow PEs (group (S)) and fast PEs (group (F)). 
At the $t$-th update attempt, the $k$-th processor belongs to group (S) 
if its local time ${\tau_k} (t)$ is less then or equal to the mean 
local time over all processors at the $t$-step. Otherwise, it belongs 
to group (F). One can define the
variance $w^2$ and the width $w_a$ for each group as follows:
%%%%%%%%%%%%%%%%%%%%%%%%%%%%%%%%%%%%%%%%%%%%%%%%%%%%%%%%%
\begin{equation}
\label{ak1}
w_{(X)}^2 (t) = \frac{1}{{L_{(X)}}(t)} \sum_{k=1}^{L_{(X)}(t)}
\left( {\tau_{k(X)}} (t) - \bar{\tau} (t) \right)^2,
\end{equation}
%%%%%%%%%%%%%%%%%%%%%%%%%%%%%%%%%%%%%%%%%%%%%%%%%%%%%%%%%
\begin{equation}
\label{ak2}
w_{a(X)} (t) = \frac{1}{{L_{(X)}}(t)} \sum_{k=1}^{L_{(X)}(t)}
\left| {\tau_{k(X)}} (t) - \bar{\tau} (t) \right|,
\end{equation}
%%%%%%%%%%%%%%%%%%%%%%%%%%%%%%%%%%%%%%%%%%%%%%%%%%%%%%%%%
where ``X" stands for either ``S" or ``F", and $L=L_{(S)}(t)+L_{(F)}(t)$. 
The variance $w^2$ and width $w_a$ of the STH can be expressed as the 
convex linear combinations:
%%%%%%%%%%%%%%%%%%%%%%%%%%%%%%%%%%%%%%%%%%%%%%%%%%%%%%%%%
\begin{equation}
\label{convex1}
w^2(t)= f_{(S)}(t) w_{(S)}^2 (t) + f_{(F)}(t) w_{(F)}^2 (t),
\end{equation}
%%%%%%%%%%%%%%%%%%%%%%%%%%%%%%%%%%%%%%%%%%%%%%%%%%%%%%%%%
\begin{equation}
\label{convex2}
w_a(t)= f_{(S)}(t) w_{a(S)} (t) + f_{(F)}(t) w_{a(F)} (t),
\end{equation}
%%%%%%%%%%%%%%%%%%%%%%%%%%%%%%%%%%%%%%%%%%%%%%%%%%%%%%%%%
where $1=f_{(S)}(t)+f_{(F)}(t)$, $0 \le f_{(S)}, f_{(F)} \le 1$. 
Explicitly, $w^2$ and $w_a$ form a {\it 1-d} simplex with vertices at 
S and F. The coefficients $f_{(S)}$  and $f_{(F)}$ are the fractions of slow 
and fast processors, respectively, in the system at each update 
attempt $t$. Figure~\ref{fig10} shows the time evolution of the 
surface widths $w_{a(S)}$, $w_{a(F)}$ and $w_a$ (Fig.~\ref{fig10}a) 
and the corresponding fractional contributions $f_{(S)}$ and $f_{(F)}$ 
(Fig.~\ref{fig10}b) for the first $500$ simulation steps. 
Quantitatively, the double peak in ${w_a} (t)$ (at about $t=10$) 
presents the weighted sum of $w_{a(S)}$ and $w_{a(F)}$ in 
accordance with Eq.~(\ref{convex2}), which is evident by matching 
the width contributions (Fig.~\ref{fig10}a) with the 
corresponding fractional contributions (Fig.~\ref{fig10}b) at 
each $t$-step. Qualitatively, the decrease in surface widths for 
$t>10$ is the effect of the constraint condition~(\ref{wrule}). In the 
particular case of simulations with $\Delta =10$ and $N_V=1000$, 
illustrated in Fig.~\ref{fig10}, initially the majority of PEs 
belongs to the slow group (about $63\%$ at $t=1$, Fig.~\ref{fig10}b), 
but as $t$ advances the STH roughens and the population of the 
slow group falls while the population in the fast group grows. As the 
population of the fast group gets larger, the fraction $u$ of PEs that 
are allowed to update falls because some of the fast PEs violate 
condition~(\ref{wrule}). While the fast PEs are waiting (i.e., no local 
time increments at the fast sites), the slow PEs are incrementing their 
local times, hence, the mean simulated time increases and therefore 
the deviation from the mean in Eq.~(\ref{ak2}) (and Eq.~
(\ref{ak1})) decreases for the fast PEs. This is the main mechanism in 
the formation of the maximum in the $w_{a(F)}$ curve. Similarly, 
the first maximum in the $w_{a(S)}$ curve is formed mainly due to the 
depopulation of the slow group. As the slow PEs are ``catching up" 
with the fast PEs within the $\Delta$-window for the update, the 
utilization $u$ increases ($20<t<50$, Fig.~\ref{fig10}b) and so do 
the widths. This secondary maximum is less pronounced  because the 
populations of the two groups are close apart. Eventually, after 
several cycles, the widths as well as the utilization reach steady 
values. 

%%%%%%%%%%%%%%%%%%%%    figure 10     %%%%%%%%%%%%%%%%%%%%%%%%%%%%
%%%%%%%%%%%%%%%%%%%%%%%%%%%%%%%%%%%%%%%%%%%%%%%%%%%%%%%%%%%%%%%%%%

In other words, the way in which the system undergoes the transition 
from the initial state to the steady state,  observed in the above 
example, is a direct consequence of the window constrained 
update scheme~(\ref{wrule}) and the particular initial condition, in 
which all PEs enter simulation with their local times equal. If this 
initial condition of the full synchronization is changed, for example, 
by assuming that at $t=0$ the local times are randomly distributed 
about some mean local time, the transition to the steady state will 
change its character. On the other hand, if at some later update 
attempt $t_s$ the system is synchronized (which is equivalent to setting 
all local simulated times to one value at $t_s$) then the 
recurrent time evolution will be repeated after $t_s$ until the 
steady state is attained. 

The above arguments can be restated in terms of the STH variance 
$w^2$ (taking Eq.~(\ref{convex1}) as the key to the analysis) 
for any system size. In the conservative PDES with a moving window 
constraint the system evolution towards a steady state follows 
essentially one path, along the above outline, for any value of 
the simulation parameters $\Delta$, $N_V$ and $L$. In our example 
we chose tentatively a very narrow $\Delta$-window and a large 
$N_V$ so the effect of the update scheme~(\ref{wrule}) is clearly 
pronounced in the evolution curves. In such a system,the 
correlations that arise due to the short-range connection 
update scheme~(\ref{crule}) are small relative to the correlations 
that arise due to the window constraint~(\ref{wrule}). Accordingly, 
in this case one can clearly deduce that a sharp fall in the 
utilization curve is the effect of a sharp population rise in the 
group of fast PEs. For example, one can read from Fig.~\ref{fig10}b 
that at $t=10$ about $25\%$ of the PEs that did not make an update 
were mostly in the group of fast PEs, so approximately about one 
half of the fast PEs updated at this update attempt. A similar 
conclusion is certainly false when each PE in the system carries a 
small number of sites (e.g., $N_V=10$) since in this case the 
correlations that originate due to the short-range connections between 
PEs may not be neglected and the utilization curve begins the fall 
at $t=0$ because the fast PEs and the slow PEs fail to satisfy 
condition~(\ref{crule}) with approximately equal frequency. Opening 
the $\Delta$-window wide (e.g., $\Delta =100$) effects the evolution 
curves in two ways. First, it slows down the build-up of the 
correlations that arise due to constraint~(\ref{wrule}). This makes 
the growth phase longer so that a transition to the steady 
state takes place at later times and over extended time intervals. 
Secondly, it softens the correlations that arise due to 
constraint~(\ref{wrule}), which smooths a transition to the steady state 
and the ripple-like features in the utilization and the width 
curves (that are clear in Fig.~\ref{fig10}) are only weakly present or 
vanish into statistical uncertainties. For example, in the worst case 
scenario of $N_V=1$ and $\Delta =100$, the time evolution towards 
the steady state follows the pattern typical for the KPZ universality 
class ($\Delta =\infty$), which suggests that the main correlation 
mechanism results from the update scheme~(\ref{crule}) in this case. 
Nonetheless, unlike the KPZ surfaces, the presence of the 
$\Delta $-window prohibits the steady-state surface width to grow 
infinitely as the system gets larger. 

\section{DISCUSSION \label{discuss}}

Our statistical analysis of the growing virtual time interface in 
conservative asynchronous PDES with a moving window constraint, shows 
that in the steady state the average utilization remains finite (and 
non zero) and scales with the system size. Similarly, the statistical 
spread of the STH remains finite and scales with the system size. 
This was demonstrated for a range of volume elements from $N_V=1$ to 
the RD-limit at $N_V=\infty$. The convergence of the utilization to 
a finite value and the convergence of the time interface width 
to a finite value as the number of processing elements infinitely 
increases, reflects positively on the ability to efficiently implement 
this type of PDES in applications. In other words, with this global 
$\Delta$-window constraint the simulation part as well as the 
measurement part of the algorithm are simultaneously scalable. 

The practical questions that should be addressed involve suitable 
implementations of the algorithm, possible modifications and 
generalizations that would facilitate applications by optimizing 
performance and thus maximizing efficiency. Such questions would 
likely be non-universal, and hence depend on the explicit problem 
being simulated. 

It follows from our analysis that the utilization as defined by a 
fraction of working processors, is not a sole measure of the 
efficiency. However, it is an important component of the efficiency. 
The case of PDES with the basic conservative scheme, when the 
measurement phase is not scalable, suggests other important factors 
that should be considered in efficiency studies. The second important 
component is the statistical spread of the virtual time surface as it 
is measured by its variance or by its mean absolute deviation. 
The third important element is the frequency and the effect of extreme 
fluctuations in the virtual time interface. The forth important factor 
is the average progress rate, which could be measured by the 
growth rate of the global minimum of the virtual time interface. 
An efficient algorithm should be characterized by the highest values 
of the utilization and the progress rate, while having small 
statistical spread in waiting times and should lack severe effects 
of the extreme time fluctuations. 

Applying the above recipe to conservative asynchronous PDES with 
a $\Delta$-window constraint, the results of our studies indicate that 
this kind of simulation presents a promise of becoming a good 
departure point towards the design of an efficient class of algorithms 
for asynchronous systems. The $\Delta$-window constraint not only 
eliminates the extreme fluctuations in the virtual time horizon but 
also controls the statistical spread of the STH and controls the 
average progress rate. The width of the $\Delta$-window can serve as a 
tuning parameter that, for a given volume load per processor, could be 
adjusted to optimize the utilization so as to maximize the efficiency.

In the conservative asynchronous PDES studied in this work, there is 
no condition imposed that would explicitly synchronize a system 
in the course of the simulations. The system is fully synchronized only 
initially and undergoes desynchronization over time (i.e., over many 
parallel steps). The degree of this desynchronization is strictly 
related to the roughening of the STH. As the simulations evolve, 
correlations between system components build up, which is reflected 
by changes in the morphology of the STH. There are two sources 
of correlations in the STH. The first is the nearest-neighbor 
communication rule that, if acted alone, would eventually lead to the 
steady state, where the entire system is correlated. In the case 
of one volume element per PE, the time to the global correlation 
is on the order of $L^{3/2}$. However, the presence of this global 
correlation does not cause an implicit synchronization nor 
does it lead to a state of near-synchronization. On the contrary, 
despite this correlation there are no global bounds on the roughening 
of the virtual time horizon: the larger the system, the more 
desynchronized it becomes over time. The nearest-neighbor communication 
rule is the essence of the conservative scheme because it ensures 
that causality is not violated in any update. The second source of 
correlations is the constraint in the form of the moving window 
condition. The moving window condition, if acted alone, would lead to 
the steady state, where the entire system is not only correlated but, 
also, is synchronized to some extent. The extent to which the system 
may become synchronized depends on the width of the moving window --- 
the roughening of the virtual time horizon is constrained to the 
$\Delta$-window width. Notice, the moving window condition is not 
necessary for the conservative scheme. Its role is to ensure that 
infinitely large desynchronization will not happen. In this sense the 
constraint condition can be seen as an implicit synchronization 
protocol. In the constrained conservative PDES, the above two 
correlation mechanisms act together: the nearest neighbor connection 
rule explicitly guarantees causality and the constraint rule implicitly 
guarantees a near-synchronization in an arbitrarily long sequence of 
update attempts.

\section{SUMMARY AND OUTLOOK \label{summary}}

We considered the conservative parallel discrete event simulations 
with the moving window constraint and studied the time evolution of 
the utilization as well as the time evolution of the stochastic time 
horizon by varying the system size (i.e., the number $L$ of processing 
elements and the number $N_V$ of sites per processing element) 
and by varying the width of the moving window. The results of our 
simulations indicate that this particular class of algorithms with 
the conservative update scheme generally scales with the system 
size. The utilization reaches a steady state value after a finite 
number of simultaneously performed parallel steps and approaches a 
finite non-zero value in the limit of infinite system size. This 
demonstrates that the simulation part of the algorithm is scalable. The 
statistical spread of the stochastic time horizon is bounded by the 
size of the moving window constraint in the limit of the infinite 
system size, which shows that the measurement part of the algorithm 
is scalable. In particular, in the limit of a large number of sites 
per processing element the results of the simulations approach 
the constrained random deposition model, which is characterized 
by a high value of utilization while permitting effective data 
management. The simultaneous scalability of both phases of 
the algorithm is an important finding because it establishes 
a solid ground for the design of new class of efficient algorithms 
for parallel processing to model the evolution of spatially extended 
interacting systems with asynchronous dynamics. Further studies are 
required in the search for possible optimal implementations. 
For example, explicitly taking into account the time required 
to find the global minimum of the STH at each step.

Aside from practical aspects of the constrained parallel conservative 
discrete event simulations that are oriented to direct 
applications such as the scalability issues, there are several 
interesting physics questions that arise in connection with the 
stochastic time surface growth. These include the development of the 
lateral correlations and transient relaxation processes. We leave 
these questions open to possible future investigations.

\begin{acknowledgments}
This work is supported by the NSF grant DMR-0113049, by the Department
of Physics and Astronomy at MSU and the ERC at
MSU. G.K. acknowledges the support of the Research Corporation through
RI0761. This research used resources of the National Energy Research
Scientific Computing Center, which is supported by the Office of
Science of the US Department of Energy under contract No.
DE-AC03-76SF00098.
\end{acknowledgments}

\appendix*
\section{The utilization data}

\begin{figure}[tp!]
\includegraphics[width=8.0cm]{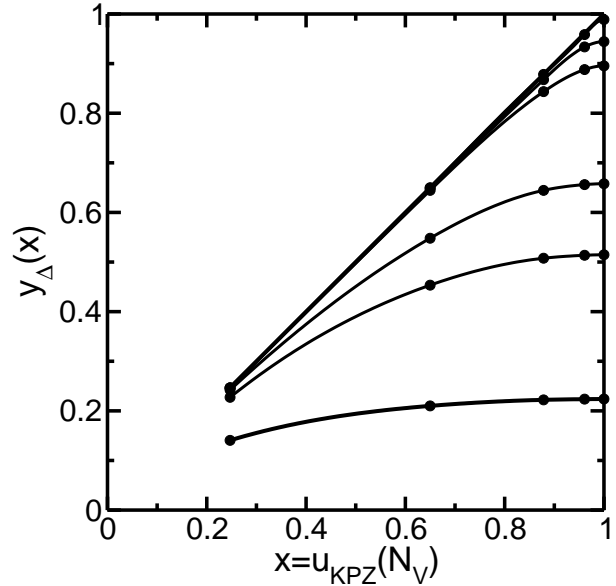}
\caption{\label{fig11} Family of utilization curves $y_{\Delta}(x)$ vs
$x=u_{KPZ}(N_V)$, illustrating the underlying idea of the fit.
For $\Delta_1 < \Delta_2< \dots < \Delta=\infty$,
$y_{\Delta_1} < y_{\Delta_2} < \dots < y_{\infty}=x$.
For $\Delta =0$, $y_{\Delta}(x)=0$ (not shown). Symbols mark
the simulation data. The cubic spline curves are guides for the eyes.}
\end{figure}

In the infinite $L$-limit the utilization is a two parameter family 
of curves (Fig.~\ref{fig6}). The two limiting curves, $u_{RD}(\Delta)$ 
and $u_{KPZ}(N_V)$, approach $u=1$ in the infinite limit of their 
arguments. One can consider either $u_{RD}$ or $u_{KPZ}$ as an 
independent variable $x$ and express the utilization as $y=y(x)$. We 
chose parameterization by $N_V$, where $x=x(N_V)=u_{KPZ}(N_V)$ 
and $y_{\Delta}(x)=y_{\Delta} \left( x(N_V) \right)$. Figure~\ref{fig11} illustrates 
the idea by plotting the utilization $y_{\Delta}(x)$ for several values 
of $\Delta$. The curves in Fig.~\ref{fig11} are a family of roots that, 
in first approximation, could be expressed by 
$y_{\Delta} (x)=a(\Delta)x^{p(\Delta)}$, where $a(\Delta)$ and 
$p(\Delta)$ have fractional values. To find $a(\Delta)$ 
and $p(\Delta)$ each curve is fitted to ``the best" two-point formula. 
Then, sequences $a(\Delta)$ and $p(\Delta)$ are expressed by fit 
functions.

%%%%%%%%%%%%%%%%%%%%    figure 11     %%%%%%%%%%%%%%%%%%%%%%%%%%%%
%%%%%%%%%%%%%%%%%%%%%%%%%%%%%%%%%%%%%%%%%%%%%%%%%%%%%%%%%%%%%%%%%%

A fit to $a(\Delta)$ is chosen in such a way that $a(0)=0$ and 
$a(\infty)=1$. In Fig.~\ref{fig11}, $a(0)=0$ corresponds to 
$y(x)=0$ because $\Delta =0$ yields $u \equiv 0$ for $L=\infty$. 
Condition $a(\infty)=1$ corresponds to $y(x)=x$ because 
$\Delta =\infty$ means $y=u_{KPZ}$. Considering the limit behavior 
of $y_{\Delta}(x)$, when $x(N_V=\infty)=1$ the coefficient 
$a(\Delta)$ must be interpreted as $u_{RD}(\Delta)$. This is also 
consistent with the alternative parameterization, where 
$x=u_{RD}(\Delta)$. Therefore, we directly identify $a(\Delta)$ with 
the approximate expression for $u_{RD}(\Delta)$. A four-point fit can 
be found as:
%%%%%%%%%%%%%%%%%%%%%%%%%%%%%%%%%%%%%%%%%%%%%%%%%%%%%%%%%
\begin{equation}
\label{appen1}
u_{RD} (\Delta) \approx a(\Delta) \cong \frac{1}
{\displaystyle 1+\frac{c_3}{\Delta ^{e_3}} -\frac{c_4}{\Delta ^{e_4}}}.
\end{equation}
%%%%%%%%%%%%%%%%%%%%%%%%%%%%%%%%%%%%%%%%%%%%%%%%%%%%%%%%%
When $c_3=15.8$, $e_3=1.07$, $c_4=12.3$ and $e_4=1.18$, 
fit~(\ref{appen1}) is good within $\pm 2\%$ relative error in the range 
$0 \le \Delta < \infty$. A simple two-point fit with $c_3=3.47$, 
$e_3=0.84$ and $c_4=e_4=0$ approximates our simulated data 
within $\pm 2.5\%$ relative difference (Fig.~\ref{fig6}, utilization 
values for $N_V=10^8$).

Considering the limits $u_{KPZ}(N_V=1) \approx 1/4$ and 
$u_{KPZ}(N_V=\infty)=1$, a four-point fit to $x(N_V)$ is:
%%%%%%%%%%%%%%%%%%%%%%%%%%%%%%%%%%%%%%%%%%%%%%%%%%%%%%%%%
\begin{equation}
\label{appen2}
u_{KPZ} (N_V) = x(N_V) \cong \frac{1}
{\displaystyle 1+\frac{c_1}{N_V ^{e_1}} +\frac{c_2}{N_V ^{e_2}}}.
\end{equation}
%%%%%%%%%%%%%%%%%%%%%%%%%%%%%%%%%%%%%%%%%%%%%%%%%%%%%%%%%
When $c_1=2.3$, $e_1=0.96$, $c_2=0.74$ and $e_2=0.4$, 
fit~(\ref{appen2}) is good to within $\pm 2\%$ relative error in the 
range $1 \le N_V < \infty$. A simple two-point fit with $c_1=3.0$, 
$e_1=0.715$ and $c_2=e_2=0$ approximates our simulated data within 
$\pm 2.5\%$ relative difference (Fig.~\ref{fig6}, utilization 
values for $\Delta = \infty$).

In fitting the power $p(\Delta)$, the limits are $p(\Delta = \infty)=1$ 
and $p(0)=0$. In Fig.~\ref{fig11}, condition $p(\Delta = \infty)=1$ 
means $y(x)=x$. Condition $p(0)=0$ expresses the fact that for small 
$\Delta$ the utilization depends mainly on $\Delta$ (not $N_V$) and, 
therefore, the exponent $p(\Delta)$ should be almost zero for small 
$\Delta$. A simple two-point formula gives 
$p(\Delta)=1/(1+2/{\Delta ^{3/4}})$. With this $p(\Delta)$, a simple 
fit to $u(N_V, \Delta) \approx a(\Delta) x(N_V) ^{p(\Delta)}$ has a 
$\pm 10\%$ relative error when $a(\Delta)$ and $x(N_V)$ are given 
by simple two-point fits. The actual power $p$ depends weakly also on 
$N_V$, $p=p(\Delta ,N_V)$. A four-point formula that accommodates 
the $N_V$ dependance can be expressed as:
%%%%%%%%%%%%%%%%%%%%%%%%%%%%%%%%%%%%%%%%%%%%%%%%%%%%%%%%%
\begin{equation}
\label{appen3}
p (\Delta, N_V) \approx \frac{1}
{\displaystyle 1+\frac{c_5(N_V)}{\Delta^{e_5(N_V)}}-
\frac{c_6(N_V)}{\Delta^{e_6(N_V)}}}.
\end{equation}
%%%%%%%%%%%%%%%%%%%%%%%%%%%%%%%%%%%%%%%%%%%%%%%%%%%%%%%%%
The fit~(\ref{fit}) is good to within $\pm 5\%$ relative uncertainty for 
all $\Delta$- and $N_V$-values when $u_{RD}$ and $u_{KPZ}$ are 
expressed by four-point fits~(\ref{appen1}) and~(\ref{appen2}), 
respectively, and when $p$ is given by~(\ref{appen3}) with the 
following fit parameters: for $N_V \ge 100$: $c_5=528.4$, $e_5=1.487$, 
$c_6=515.1$, $e_6=1.609$; for $N_V < 10$: $c_5=17.43$, $e_5=1.406$, 
$c_6=15.3$, $e_6=1.687$; otherwise: $c_5=5.345$, $e_5=0.627$, 
$c_6=0.095$, $e_6=0.045$.

\end{document}